# Gauged $N = 2$ Supergravity in Nine-Dimensions and Domain Wall Solutions

Hitoshi NISHINO[1] and Subhash RAJPOOT[2]

*Department of Physics & Astronomy*
*California State University*
*Long Beach, CA 90840*

**Abstract**

We present massive $N = 2$ supergravity with $SO(2)$-gauging in nine-dimensions by direct construction. A full lagrangian and transformation rules are fixed, respectively up to quartic and quadratic fermion terms. Corresponding to the generalized Scherk-Schwarz dimensional reduction utilizing $SL(2, \mathbb{R})$ symmetry, this theory allows three arbitrary mass parameters $m_0$, $m_1$ and $m_2$ in addition to the minimal gauge coupling $g$, so that our system has the most general form compared with other results in the past. Unlike ordinary gauged maximal supergravity theories in other dimensions, the scalar potential is positive definite for arbitrary values of the mass parameters. As an application, we also analyze the stability and supersymmetry for 7-brane domain wall solutions for this gauged maximal supergravity, keeping the three mass parameters.

[1]E-Mail: hnishino@csulb.edu

[2]E-Mail: rajpoot@csulb.edu

## 1. Introduction

There has been increasing recognition of the importance of maximal supergravity theories in dimensions between four and eleven, due to their relationship with M-theory or superstring theory. Gauged or massive maximal supergravity, such as type IIA theory in 10D [1] have more importance, due to the possible duality with non-massive theories [2]. Furthermore, these massive maximal supergravity theories are dual to other non-massive theories, such as massive type IIA theory is related to type IIB theory under T-duality [3]. Additionally, the importance of massive supergravity theories with cosmological constants is associated with the AdS background which in turn is related to brane-domain wall/AdS/CFT correspondence [4].

Massive supergravity theories are generated by new generalized dimensional reduction scheme utilizing certain $\sigma$-model symmetry [5][6], as a generalization of Scherk-Schwarz type dimensional reduction [7]. Such massive supergravity is possible in space-time dimensions up to $D \le 10$ [8], while gauged supergravity is possible only in $D \le 9$ [8]. Therefore, it seems that 9D is the unique maximal space-time dimensions for massive maximal supergravity with gauging [8].

Motivated by these developments, there have been some works on maximal supergravity in 9D. The first work was in [9], in which the non-gauged $N = 2$ supergravity lagrangian and transformation rule were given. However, the results in [9] seem to suffer from certain flaws caused by technical but fundamental mistakes. One example is the mistreatment of the barred spinors, in the dimensional reduction from 11D into 9D. For example, the barred spinor parameter of supersymmetry $\overline{\epsilon}$ in 11D does not stay just the same $\overline{\epsilon}$ in 9D any longer, after the simple dimensional reduction [7]. The reason is that the charge conjugation matrix $\widehat{C}$ in 11D is antisymmetric, while that $C$ in 9D is symmetric. Therefore, there must be an extra antisymmetric matrix, such as the second Pauli matrix $\sigma_2$ should be present: $\widehat{C} = \sigma_2 \otimes C$. This affects many terms given in [9], such as the missing $\sigma_2$-matrix for the $F_{\mu\nu\rho\sigma}$-terms in the gravitino transformation rule $\delta_{\rm Q}\psi_\mu$, that in turn leads to the non-closure of supersymmetry on the neunbein $e_\mu{}^m$, caused by the unwanted $e_\mu{}^m(\overline{\epsilon}_2\gamma^{\nu\rho\sigma\tau}\epsilon_1)F_{\nu\rho\sigma\tau}$-term. This is due to the flipping property of gamma-matrices in 9D that leaves this combination non-vanishing, thus violating the fundamental closure of supersymmetry.

The first version of $SO(2)$-gauging of maximal supergravity in 9D was given in [10], based on a generalized dimensional reduction combining the $SL(2,\mathbb{R})$ symmetry with the extra coordinate dependence [5][6] from $N = 2$ supergravity 10D into 9D, as a generalization of Scherk-Schwarz dimensional reduction [7]. Recently, it has been claimed [11] that the Minkowski background in 9D $(\text{Mink})_9$ is realized as the vacuum solution for gauged $N = 2$ supergravity, without preserving supersymmetry. However, this is unusual for supergravity, because in ordinary supergravity a scalar potential is minimized with supersymmetry, unless there is a nonzero cosmological constant. On the other hand, it has been already well-known that the scalar potential for gauged $N = 2$ supergravity in 9D is positive definite [10][12]. To deal with such a subtle issue, it seems also imperative to derive both the total lagrangian and the transformation rule in a consistent manner. For example, even though supersymmetry transformation rules for gauged maximal supergravity in 9D were given in [11], it is still important to see the mutual consistency between the total lagrangian and transformation rules.

Considering these recent developments, it seems imperative to establish a more complete system of massive gauged $N = 2$ supergravity in 9D with a consistent lagrangian and supersymmetry transformation rule. In this Letter, we give a more complete result for $N = 2$ gauged maximal supergravity in 9D, based on direct construction within 9D, instead of dimensional reduction from 11D or 10D. We perform the $SO(2)$-gauging of the $\sigma$-model coset $SL(2, \mathbb{R})/SO(2)$, by a vector field available in the multiplet. We keep all the three possible mass parameters $m_1$, $m_2$ and $m_3$ in addition to the minimal gauge coupling $g$. These mass parameters corresponds to the generalization [5] of the dimensional reduction by [7], so that our system is the most general compared with other systems [9][10][11] in the past.

## 2. Lagrangian and Transformation Rule

The field content of our gauged $N = 2$ supergravity in 9D is $(e_\mu{}^m, \psi_{\mu\, A}, A_{[3]}, B_{[2]\,\alpha}, A_\mu, B_{\mu\,\alpha}, \chi_{i\,A}, \lambda_A, L_\alpha{}^i, \varphi)$. Here the gravitino $\psi_{\mu\, A}$ is a pair of Majorana spinors forming $\mathbf{2}_{\rm s}$ of $SO(2)$ with the spinorial $\mathbf{2}_{\rm s}$-indices $A, B, \cdots$ usually suppressed. The $A_{\mu\nu\rho}$ and $A_\mu$ are both real, while $B_{\mu\nu\,\alpha}$ and $B_{\mu\nu\rho\,\alpha}$ both carry the curved two-dimensional index $\alpha = 1, 2$ for the coset $SL(2, \mathbb{R})/SO(2)$. The fermion $\chi_{i\,A}$ carries both the $\mathbf{2}_{\rm v}$-index $i$ of $SO(2)$ and $\mathbf{2}_{\rm s}$-index $A$ of $SO(2)$, the latter of which is usually suppressed. This $\chi_i$ is also subject to the extra constraint $\tau_i \chi_i = 0$ in order to have the right degrees of freedom. The fermion $\lambda_A$ is in the $\mathbf{2}_{\rm s}$ of $SO(2)$. The real scalars $L_\alpha{}^i$ are the coset representatives for $SL(2, \mathbb{R})/SO(2)$, while $\varphi$ is a real scalar dilaton. As for notations, we use $(\eta_{mn}) \equiv \text{diag.}\,(+ - - \cdots -)$, $\{\gamma_m, \gamma_n\} = +2\eta_{mn} I$, and $\epsilon^{[9-n][n]}\gamma_{[n]} = (-1)^{n(n-1)/2}\,(n!)\,\gamma^{[9-n]}$. Here we use the symbol such as $[n]$ for a normalized antisymmetrized indices, *e.g.,* $\gamma^{[n]}$ is equivalent to $\gamma^{m_1 \cdots m_n}$. We use also the $\tau$-matrices defined by

$$\tau_1 \equiv \begin{pmatrix} 0 & 1 \\ 1 & 0 \end{pmatrix} \quad , \quad \tau_2 \equiv \begin{pmatrix} 1 & 0 \\ 0 & -1 \end{pmatrix} \quad , \quad \tau_3 \equiv \begin{pmatrix} 0 & -1 \\ 1 & 0 \end{pmatrix} \quad , \tag{2.1}$$

so that $\tau_i \tau_j = \delta_{ij} + \tau_{ij} = \delta_{ij} + \epsilon_{ij} \tau_3$, and $(\tau_3)^2 = -I$. We need to assign the indices $i = 1, 2$ for $\mathbf{2}_{\rm v}$ on the *symmetric* $2 \times 2$ matrices, and this is why our $\tau$-matrices have the numbering different from the standard Pauli matrices. The multiplication of fermions by these matrices are such as $(\overline{\psi}_\mu \tau_3 \lambda) \equiv \overline{\psi}_{mA} (\tau_3)_{AB} \lambda_B$, where due to the positive definite metric for $SO(2)$, there is no distinction for super/subscripts of $A, B$. For each value of the $A$-index, the flipping property for the spinorial product is such that $(\overline{\psi}\gamma^{[n]}\chi) = -(-1)^{n(n-1)/2}(\overline{\chi}\gamma^{[n]}\psi)$, as in the $N = 1$ supergravity in 9D [13].[3] Relevantly, we have $(i\overline{\psi}\gamma^{[n]}\chi)^\dagger = +(i\overline{\psi}\gamma^{[n]}\chi)$.

[3]There is a crucial sign error with a flipping property equation in [13]. The equation

The coset representative form the $\sigma$-model quantities $\mathcal{P}$ and $\mathcal{Q}$ defined by [8]

$$
\begin{aligned}
& X_{\mu\, ij} \equiv L_i{}^\alpha (\partial_\mu L_{\alpha j} + g \eta_{\alpha\beta} \epsilon^{\beta\gamma} A_\mu L_{\gamma j}) \equiv L_i{}^\alpha \mathcal{D}_\mu L_{\alpha j} = \mathcal{P}_{\mu\, ij} + \mathcal{Q}_{\mu\, ij} \quad , \\
& \mathcal{P}_{\mu ij} \equiv X_{\mu(ij)} \quad , \qquad \mathcal{Q}_{\mu ij} \equiv X_{\mu[ij]} \equiv \mathcal{Q}_\mu \, \epsilon_{ij} \quad , \qquad \mathcal{P}_{\mu ii} \equiv 0 \quad , \\
& \eta_{\alpha\beta} \equiv m_0 \delta_{\alpha\beta} + m_1 (\tau_1)_{\alpha\beta} + m_2 (\tau_2)_{\alpha\beta} = \eta_{\beta\alpha} \quad .
\end{aligned} \tag{2.2}
$$

The condition $\mathcal{P}_{\mu ii} \equiv 0$ is from the unimodular nature of $L_\alpha{}^i$ for $SL(2,\mathbb{R})/SO(2)$. The $\eta_{\alpha\beta}$-matrix can be also regarded as a 'metric' tensor, in the sense that it lowers the index $\beta$ on $\epsilon^{\beta\gamma}$ in $X_{\mu\, ij}$. Note that the three constants $m_0$, $m_1$ and $m_2$ correspond to the mass parameters in [10] created by a generalized dimensional reduction [5] from 10D into 9D using the $SL(2,\mathbb{R})$-symmetry, as a generalization of the Scherk-Schwarz dimensional reduction [7]. The explicit representation of the coset representatives $L$'s and the metric on the coset $SL(2,\mathbb{R})/SO(2)$ [10][11] are

$$
\begin{aligned}
& (L_{\alpha i}) = e^{\phi/2} \begin{pmatrix} e^{-\phi} & -\rho \\ 0 & -1 \end{pmatrix} \quad , \qquad (L_i{}^\alpha) = e^{-\phi/2} \begin{pmatrix} e^{\phi} & -e^{\phi}\rho \\ 0 & -1 \end{pmatrix} \quad , \\
& (g_{\alpha\beta}) \equiv (L_{\alpha i} L_{\beta i}) = e^{\phi} \begin{pmatrix} \rho^2 + e^{-2\phi} & \rho \\ \rho & 1 \end{pmatrix} \quad , \quad (g^{\alpha\beta}) \equiv (L_i{}^\alpha L_i{}^\beta) = e^{\phi} \begin{pmatrix} 1 & -\rho \\ -\rho & \rho^2 + e^{-2\phi} \end{pmatrix} ,
\end{aligned} \tag{2.3}
$$

where $g_{\alpha\beta}$ corresponds to the matrix $\mathcal{M}$ in [10].

Another important quantity for gauged supergravity is the matrix $S_{ij}$ defined by

$$
\begin{aligned}
S_{ij} & \equiv \eta_{\alpha\beta} L_i{}^\alpha L_j{}^\beta = S_{ji} \\
& = \begin{pmatrix} (m_0 + m_2) e^{\phi} - 2 m_1 e^{\phi} \rho + (m_0 - m_2) e^{\phi} \rho^2 & -m_1 + (m_0 - m_2)\rho \\ -m_1 + (m_0 - m_2)\rho & (m_0 - m_2) e^{-\phi} \end{pmatrix} \equiv \begin{pmatrix} A & B \\ B & C \end{pmatrix} , \\
S & \equiv S_{ii} = \eta_{\alpha\beta} g^{\alpha\beta} = A + C \quad ,
\end{aligned} \tag{2.4}
$$

which is very similar to other maximally extended supergravity models [8], such as $N = 2$ in 8D [15] or $N = 4$ in 7D [16].

With these preliminaries, we are ready to give our lagrangian

$$
\begin{aligned}
e^{-1} \mathcal{L} = & - \tfrac{1}{4} R - \tfrac{i}{2} (\overline{\psi}_\mu \gamma^{\mu\nu\rho} \mathcal{D}_\nu \psi_\rho) - \tfrac{1}{48} e^{4\varphi/\sqrt{7}} F_{[4]}^2 + \tfrac{1}{12} e^{-2\varphi/\sqrt{7}} (G_{[3]\, i})^2 - \tfrac{1}{4} e^{-8\varphi/\sqrt{7}} F_{\mu\nu}^2 \\
& - \tfrac{1}{4} e^{6\varphi/\sqrt{7}} (G_{\mu\nu\, i})^2 + \tfrac{1}{2} (\partial_\mu \varphi)^2 + \tfrac{i}{2} (\overline{\chi}_i \gamma^\mu \mathcal{D}_\mu \chi_i) + \tfrac{i}{2} (\overline{\lambda} \gamma^\mu \mathcal{D}_\mu \lambda) + \tfrac{1}{4} (\mathcal{P}_{\mu\, ij})^2 \\
& + e^{2\varphi/\sqrt{7}} F_{[4]} \Big[ - \tfrac{i}{96} (\overline{\psi}^\rho \tau_3 \gamma_{[\rho|} \gamma^{\mu\nu [4]} \gamma_{|\sigma]} \psi^\sigma) - \tfrac{i}{24\sqrt{14}} (\overline{\psi}_\mu \tau_3 \gamma^{[4]} \gamma^\mu \lambda) \\
& \qquad - \tfrac{i}{96} (\overline{\chi}_i \tau_3 \gamma^{[4]} \chi_i) + \tfrac{i}{224} (\overline{\lambda} \tau_3 \gamma^{[4]} \lambda) \Big]
\end{aligned}
$$

$\left(\overline{\psi}\gamma^{\mu_1\cdots\mu_n}\chi\right) = (-1)^n \left(\overline{\chi}\gamma^{\mu_n\cdots\mu_1}\psi\right)$ given below eq. (22b) in [13] should be replaced by $\left(\overline{\psi}\gamma^{\mu_1\cdots\mu_n}\chi\right) = -\left(\overline{\chi}\gamma^{\mu_n\cdots\mu_1}\psi\right) = -(-1)^{n(n-1)/2}\left(\overline{\chi}\gamma^{\mu_1\cdots\mu_n}\psi\right)$. This can be reconfirmed by the aid of [14].

$$
\begin{aligned}
&+ e^{-\varphi/\sqrt{7}} G_{[3]\,i} \Big[ - \tfrac{i}{24} (\overline{\psi}^{\rho} \tau_i \gamma_{[\rho|} \gamma^{[3]} \gamma_{|\sigma]} \psi^{\sigma}) + \tfrac{i}{12} (\overline{\psi}_{\mu} \gamma^{[3]} \gamma^{\mu} \chi_i) \\
&\qquad\qquad + \tfrac{i}{12\sqrt{4}} (\overline{\psi}_{\mu} \tau_i \gamma^{[3]} \gamma^{\mu} \lambda) G_{[3]\,i} + \tfrac{i}{6\sqrt{14}} (\overline{\chi}_i \gamma^{[3]} \lambda) - \tfrac{i}{28} (\overline{\lambda} \tau_i \gamma^{[3]} \lambda) \Big] \\
&+ e^{-4\varphi/\sqrt{7}} F_{\mu\nu} \Big[ - \tfrac{i}{8} (\overline{\psi}^{\rho} \gamma_{[\rho|} \gamma^{\mu\nu} \gamma_{|\sigma]} \psi^{\sigma}) + \tfrac{i}{\sqrt{14}} (\overline{\psi}_{\mu} \gamma^{\rho\sigma} \gamma^{\mu} \lambda) \\
&\qquad\qquad + \tfrac{i}{8} (\overline{\chi}_i \gamma^{\mu\nu} \chi_i) - \tfrac{9i}{56} e^{-4\varphi/\sqrt{7}} (\overline{\lambda} \gamma^{\mu\nu} \lambda) \Big] \\
&+ e^{3\varphi/\sqrt{7}} G_{\mu\nu\,i} \Big[ - \tfrac{i}{8} (\overline{\psi}^{\rho} \tau_i \gamma_{[\rho|} \gamma^{\mu\nu} \gamma_{|\sigma]} \psi^{\sigma}) + \tfrac{i}{4} (\overline{\psi}_{\mu} \gamma^{\rho\sigma} \gamma^{\mu} \chi_i) - \tfrac{3i}{4\sqrt{14}} (\overline{\psi}_{\mu} \tau_i \gamma^{\rho\sigma} \gamma^{\mu} \lambda) \\
&\qquad\qquad + \tfrac{3i}{2\sqrt{14}} (\overline{\chi}_i \gamma^{\mu\nu} \lambda) - \tfrac{i}{28} e^{3\varphi/\sqrt{7}} (\overline{\lambda} \tau_i \gamma^{\mu\nu} \lambda) \Big] \\
&+ \tfrac{i}{2} (\overline{\psi}_{\mu} \tau_i \gamma^{\nu} \gamma^{\mu} \chi_j) \mathcal{P}_{\nu\,ij} + \tfrac{i}{\sqrt{2}} (\overline{\psi}_{\mu} \gamma^{\nu} \gamma^{\mu} \lambda) \partial_{\nu} \varphi \\
&+ \tfrac{1}{576} e^{-1} \epsilon^{[4]\,[4]'\mu} F_{[4]} F_{[4]'} A_{\mu} - \tfrac{1}{216} e^{-1} \epsilon^{[3]\,[3]'\,[3]''}\, \epsilon^{\alpha\beta}\, \widetilde{G}_{[3]\,\alpha} \widetilde{G}_{[3]'\,\beta} A_{[3]''} \\
&- \tfrac{1}{36} e^{-1} \epsilon^{[3]\,[3]'\,[2]\mu} \epsilon^{\alpha\beta} \epsilon^{\gamma\delta} \widetilde{G}_{[3]\,\alpha} \widetilde{G}_{[3]'\,\beta} B_{[2]\,\gamma} B_{\mu\,\delta} \\
&- \tfrac{i}{16} g e^{4\varphi/\sqrt{7}} S (\overline{\psi}_{\mu} \tau_3 \gamma^{\mu\nu} \psi_{\nu}) - \tfrac{i}{2\sqrt{14}} g e^{4\varphi/\sqrt{7}} S (\overline{\psi}_{\mu} \tau_3 \gamma^{\mu} \lambda) - \tfrac{i}{16} g e^{4\varphi/\sqrt{7}} S (\overline{\chi}_i \tau_3 \chi_i) \\
&- \tfrac{9i}{116} g e^{4\varphi/\sqrt{7}} S (\overline{\lambda} \tau_3 \lambda) + \tfrac{i}{2} g e^{4\varphi/\sqrt{7}} S_{ij} (\overline{\chi}_i \tau_3 \chi_j) + \tfrac{i}{4} g e^{4\varphi/\sqrt{7}} S_{ij} (\overline{\psi}_{\mu} \tau_3 \tau_i \gamma^{\mu} \chi_j) \\
&- \tfrac{2i}{\sqrt{14}} S_{ij} (\overline{\chi}_i \tau_3 \tau_j \lambda) - \tfrac{1}{32} g^2 e^{8\varphi/\sqrt{7}} \Big[ 2 (S_{ij})^2 - S^2 \Big] \quad , \qquad (2.5)
\end{aligned}
$$

up to quartic fermion terms. Our action $I = \int d^9 x\, \mathcal{L}$ is invariant under supersymmetry

$$
\begin{aligned}
\delta_Q e_{\mu}{}^{m} = &- i (\overline{\epsilon} \gamma^m \psi_{\mu}) \quad , \\
\delta_Q \psi_{\mu} = &+ \mathcal{D}_{\mu} \epsilon + \tfrac{1}{112} e^{2\varphi/\sqrt{7}} \tau_3 (\gamma_{\mu}{}^{\nu[3]} - \tfrac{16}{3} \delta_{\mu}{}^{\nu} \gamma^{[3]}) \epsilon \widehat{F}_{\nu[3]} \\
&+ \tfrac{1}{42} e^{-\varphi/\sqrt{7}} \tau_i (\gamma_{\mu}{}^{\nu[2]} - \tfrac{15}{2} \delta_{\mu}{}^{\nu} \gamma^{[2]}) \widehat{G}_{\nu[2]\,i} + \tfrac{1}{28} e^{-4\varphi/\sqrt{7}} (\delta_{\mu}{}^{\nu\rho} - 12 \delta_{\mu}{}^{\nu} \gamma^{\rho}) \epsilon \widehat{F}_{\nu\rho} \\
&+ \tfrac{1}{28} e^{3\varphi/\sqrt{7}} \tau_i (\gamma_{\mu}{}^{\nu\rho} - 12 \delta_{\mu}{}^{\nu} \gamma^{\rho}) \epsilon \widehat{G}_{\nu\rho\,i} - \tfrac{1}{56} g e^{4\varphi/\sqrt{7}} \tau_3 \gamma_{\mu} \epsilon S \equiv \widehat{\mathcal{D}}_{\mu} \epsilon \quad , \\
\delta_Q A_{\mu\nu\rho} = &- \tfrac{3i}{2} e^{-2\varphi/\sqrt{7}} (\overline{\epsilon} \tau_3 \gamma_{[\mu\nu} \psi_{\rho]}) + \tfrac{i}{\sqrt{14}} e^{-2\varphi/\sqrt{7}} (\overline{\epsilon} \tau_3 \gamma_{\mu\nu} \lambda) \\
&+ 6 \epsilon^{\alpha\beta} B_{[\mu\nu|\,\alpha} (\delta_Q B_{|\nu]\,\beta}) + 24 \epsilon^{\alpha\beta} A_{[\mu|} B_{|\nu|\,\alpha} (\delta_Q B_{|\rho]\,\beta}) \\
\delta_Q B_{\mu\nu\,\alpha} = &- i e^{\varphi/\sqrt{7}} L_{\alpha}{}^{i} (\overline{\epsilon} \tau_i \gamma_{[\mu} \psi_{\nu]}) - \tfrac{i}{2} e^{\varphi/\sqrt{7}} L_{\alpha}{}^{i} (\overline{\epsilon} \gamma_{\mu\nu} \chi_i) \\
&- \tfrac{i}{2\sqrt{14}} L_{\alpha}{}^{i} (\overline{\epsilon} \tau_i \gamma_{\mu\nu} \lambda) - 4 B_{[\mu|\,\alpha} (\delta_Q A_{|\nu]}) \quad , \\
\delta_Q A_{\mu} = &- \tfrac{i}{2} e^{4\varphi/\sqrt{7}} (\overline{\epsilon} \psi_{\mu}) - \tfrac{2i}{\sqrt{14}} e^{4\varphi/\sqrt{7}} (\overline{\epsilon} \gamma_{\mu} \lambda) \quad , \\
\delta_Q B_{\mu\,\alpha} = &- \tfrac{i}{2} e^{-3\varphi/\sqrt{7}} L_{\alpha\,i} (\overline{\epsilon} \tau_i \psi_{\mu}) - \tfrac{i}{2} e^{-3\varphi/\sqrt{7}} L_{\alpha\,i} (\overline{\epsilon} \gamma_{\mu} \chi_i) + \tfrac{3i}{2\sqrt{14}} L_{\alpha\,i} (\overline{\epsilon} \tau_i \gamma_{\mu} \lambda) \quad , \\
\delta_Q \chi_i = &+ \tfrac{1}{2} \tau_j \gamma^{\mu} \epsilon \widehat{\mathcal{P}}_{\mu\,ij} - \tfrac{1}{4} e^{3\varphi/\sqrt{7}} (\delta_{ij} - \tfrac{1}{2} \tau_i \tau_j) \gamma^{\mu\nu} \epsilon \widehat{G}_{\mu\nu\,j} \\
&- \tfrac{1}{8} e^{-\varphi/\sqrt{7}} (\delta_{ij} - \tfrac{1}{2} \tau_i \tau_j) \gamma^{[3]} \epsilon \widehat{G}_{[3]\,j} + \tfrac{1}{8} g e^{4\varphi/\sqrt{7}} (\delta_{ij} - \epsilon_{ij} \tau_3) \tau_3 \tau_k \epsilon S_{jk} \quad , \\
\delta_Q \lambda = &- \tfrac{1}{\sqrt{14}} e^{-4\varphi/\sqrt{7}} \gamma^{\mu\nu} \epsilon \widehat{F}_{\mu\nu} + \tfrac{3}{4\sqrt{14}} e^{3\varphi/\sqrt{7}} \tau_i \gamma^{\mu\nu} \epsilon \widehat{G}_{\mu\nu\,i}
\end{aligned}
$$

$$- \tfrac{1}{12\sqrt{14}} e^{-\varphi/\sqrt{7}} \tau_i \gamma^{[3]} \epsilon \widehat{G}_{[3]\, i} + \tfrac{1}{24\sqrt{14}} e^{2\varphi/\sqrt{7}} \tau_3 \gamma^{[4]} \epsilon \widehat{F}_{[4]} + \tfrac{1}{\sqrt{2}} \gamma^\mu \epsilon \widehat{D}_\mu \varphi$$
$$+ \tfrac{1}{2\sqrt{14}} g e^{4\varphi/\sqrt{7}} \tau_3 \epsilon S \ ,$$
$$L_i{}^\alpha \delta_Q L_{\alpha\, j} = - i(\overline{\epsilon} \tau_i \chi_j) \ , \quad \delta_Q \varphi = - \tfrac{i}{\sqrt{2}} (\overline{\epsilon} \lambda) \ , \tag{2.6}$$

up to quadratic fermion terms. As usual, the *hatted* field strengths, *e.g.,* $\widehat{G}_{\mu\nu\mu\,\alpha}$ , *etc.* are supercovariantized ones [8]. Each term in $\delta_Q \chi_i$ has the right projection, such that the extra constraint $\tau_i \chi_i = 0$ is desirably satisfied. Note that all the three mass parameters $m_0$, $m_1$ and $m_2$ are arbitrary in our formulation with the $SO(2)$-gauging, so that our result here is the most general compared with other past references [9][10][11].

The field strengths are defined by

$$F_{\mu\nu\rho\sigma} \equiv 4\mathcal{D}_{[\mu} A_{\nu\rho\sigma]} + 12 \epsilon^{\alpha\beta} B_{[\mu\nu|\,\alpha} G_{|\rho\sigma]\,\beta} - 48 \epsilon^{\alpha\beta} A_{[\mu|} B_{|\nu|\,\alpha} G_{|\rho\sigma]\,\beta}$$
$$- 24 g \epsilon^{\alpha\beta} \eta_{\beta\gamma} \epsilon^{\gamma\delta} A_{[\mu|} B_{|\nu\rho|\,\alpha} B_{|\sigma]\,\delta} + 3 g \epsilon^{\alpha\beta} \eta_{\beta\gamma} \epsilon^{\gamma\delta} B_{[\mu\nu|\,\alpha} B_{|\rho\sigma]\,\delta} \ , \tag{2.7a}$$

$$G_{\mu\nu\rho\,\alpha} \equiv 3\mathcal{D}_{[\mu} B_{\nu\rho]\,\alpha} - 6 F_{[\mu\nu} B_{\rho]\,\alpha} \ , \quad \widetilde{G}_{\mu\nu\rho\,\alpha} \equiv G_{\mu\nu\rho\,\alpha} + 6 A_{[\mu} G_{\nu\rho]\,\alpha} \ , \tag{2.7b}$$

$$G_{\mu\nu\,\alpha} \equiv 2\mathcal{D}_{[\mu} B_{\nu]\,\alpha} - \tfrac{1}{2} g \eta_{\alpha\beta} \epsilon^{\beta\gamma} B_{\mu\nu\,\gamma} \ , \quad F_{\mu\nu} \equiv 2\partial_{[\mu} A_{\nu]} \ . \tag{2.7c}$$

The $\widetilde{G}_{\mu\nu\rho\,\alpha}$ is used in the Chern-Simons terms in the lagrangian. All the field strengths with the index $i$ are defined by the multiplication of the above field strengths by $L_i{}^\alpha$ *e.g.,* $G_{\mu\nu\rho\, i} \equiv L_i{}^\alpha G_{\mu\nu\rho\,\alpha}$. As usual in supergravity theories [8], the extra transformations $B_{[1]} \wedge \delta_Q A_{[1]}$ in $\delta_Q B_{[2]}$ or $B_{[2]} \wedge \delta_Q B_{[1]}$ or $A_{[1]} \wedge B_{[1]} \wedge \delta_Q B_{[1]}$ in $\delta_Q A_{[3]}$ are determined by the requirement that the supersymmetry transformations of their field strengths will have only field strengths [8]. The covariant derivative $\mathcal{D}_\mu$ acts on fermions and potentials, as

$$\mathcal{D}_{[\mu|} \psi_{|\nu]} \equiv D_{[\mu|}(\widehat{\omega}) \psi_{|\nu]} + \tfrac{1}{4} \mathcal{Q}_{[\mu|\, ij} \tau_{ij} \psi_{|\nu]} = D_{[\mu|}(\widehat{\omega}) \psi_{|\nu]} + \tfrac{1}{2} \mathcal{Q}_{[\mu|} \tau_3 \psi_{|\nu]} \ ,$$
$$\mathcal{D}_\mu \chi_i \equiv D_\mu(\widehat{\omega}) \chi_i + \tfrac{1}{4} \mathcal{Q}_{\mu\, jk} \tau_{jk} \chi_i + \mathcal{Q}_{\mu ij} \chi_j \ , \quad \mathcal{D}_\mu \lambda \equiv D_\mu(\widehat{\omega}) \lambda + \tfrac{1}{4} \mathcal{Q}_{\mu\, ij} \tau_{ij} \lambda \ ,$$
$$\mathcal{D}_\mu B_{\nu\rho\,\alpha} \equiv \partial_\mu B_{\nu\rho\,\alpha} + g \eta_{\alpha\beta} \epsilon^{\beta\gamma} A_\mu B_{\nu\rho\,\gamma} \ , \quad \mathcal{D}_\mu B_{\nu\,\alpha} \equiv \partial_\mu B_{\nu\,\alpha} + g \eta_{\alpha\beta} \epsilon^{\beta\gamma} A_\mu B_{\nu\,\gamma} \ , \tag{2.8}$$

where the $\mathcal{Q}$'s has also the implicit $SO(2)$ minimal coupling *via* (2.2). The derivatives on the $S$'s give

$$\partial_\mu S_{ij} = -2 S_{(i|k} X_{\mu\,|j)k} \ , \quad \partial_\mu S = -2 S_{ij} X_{\mu\, ij} = -2 S_{ij} \mathcal{P}_{\mu\, ij} \ . \tag{2.9}$$

As in the massive type IIA [1] or the gauged $N = 2$ in 8D [15], we need to put a linear term in the tensor field $B_{\mu\nu\,\alpha}$ in the field strength $G_{\mu\nu\,\alpha}$ for the gauged case with $g \neq 0$. Relevantly, some useful relationships for invariance confirmation are the Bianchi identities

$$\mathcal{D}_{[\mu} G_{\nu\rho]\,\alpha} \equiv -\tfrac{1}{6}\, g\, \eta_{\alpha\beta}\, \epsilon^{\beta\gamma}\, G_{\mu\nu\rho\,\gamma} \ , \quad \mathcal{D}_{[\mu} G_{\nu\rho\sigma]\,\alpha} \equiv -3 F_{[\mu\nu} G_{\rho\sigma]\,\alpha} \ , \tag{2.10a}$$

$$\mathcal{D}_{[\mu} F_{\nu\rho\sigma\tau]} = +4 \epsilon^{\alpha\beta} G_{[\mu\nu\rho|\,\alpha} G_{|\sigma\tau]\,\beta} \ . \tag{2.10b}$$

Eq. (2.10b) confirms the validity of the explicitly $g$-dependent terms in (2.7a).

Even though there are some $A_\mu$-explicit terms in (2.7a), we can show that $F_{\mu\nu\rho\sigma}$ is invariant under the local $SO(2)$ transformation with the parameter $\alpha$:

$$\delta_\alpha A_\mu = \partial_\mu \alpha \quad , \quad \delta_\alpha B_{\mu\,\alpha} = -g\alpha\eta_{\alpha\beta}\epsilon^{\beta\gamma}B_{\mu\,\gamma} \quad , \quad \delta_\alpha B_{\mu\nu\,\alpha} = -g\alpha\eta_{\alpha\beta}\epsilon^{\beta\gamma}B_{\mu\nu\,\gamma} \quad . \tag{2.11}$$

We can show also the covariance (or invariance) of all the $G$-field (or $F$-field) strengths, *except* the *tilded* one $\widetilde{G}_{\mu\nu\rho\,\alpha}$, as will be mentioned in the paragraph below (2.12).

As has been well-known for massive maximal supergravity theories [1], the $SO(2)$-gauging breaks the covariance of the field strength $G_{\mu\nu\rho\,\alpha}$ under the proper antisymmetric tensor gauge transformation of $B_{\mu\nu\,\alpha}$, upsets the right propagating degrees of freedom. However, this is compensated by the absorption of $B_{\mu\nu\,\alpha}$ into $G_{\mu\nu\,\alpha}$. After this, $B_{\mu\nu\,\alpha}$ loses its original tensor gauge covariance, but this does not pose any problem, because it becomes massive with the right propagating degrees of freedom. A similar situation can be also found for the gauged maximal supergravity in 8D [15].

Note that there are two ways to recover the non-gauged case of $N = 2$ supergravity in 9D [9],[4] either by putting $g = 0$ or by requiring all the mass parameters to zero: $m_0 = m_1 = m_2 = 0$. Even though $\eta_{\alpha\beta}$ becomes non-invertible in the latter case, it does not matter, because $\eta_{\alpha\beta}$ is always with the minimal couplings with $gA_\mu$ which are not needed in the non-gauged case anyway.

The validity of the Chern-Simons terms with the $\epsilon$-tensors in (2.5) can be reconfirmed by the supersymmetric variation: In terms of differential forms, we get

$$\begin{aligned}&\delta_Q \left[ \tfrac{1}{576} F_4 F_4 A_1 - \tfrac{1}{216}\epsilon^{\alpha\beta}\widetilde{G}_{3\alpha}\widetilde{G}_{3\beta}A_3 - \tfrac{1}{36}\epsilon^{\alpha\beta}\epsilon^{\gamma\delta}\widetilde{G}_{3\alpha}\widetilde{G}_{3\beta}B_{2\gamma}B_{1\delta}\right] \\ &= \tfrac{1}{144}(\delta_Q^{(0)}A_3)F_4F_2 + \tfrac{1}{572}F_4F_4(\delta_Q^{(0)}A_1) + \tfrac{1}{144}\epsilon^{\alpha\beta}(\delta_Q^{(0)}B_{2\alpha})G_{3\beta}F_4 - \tfrac{1}{216}\epsilon^{\alpha\beta}G_{3\alpha}G_{3\beta}(\delta_Q^{(0)}A_3)\ ,\end{aligned} \tag{2.12}$$

where $\delta_Q^{(0)}$ is for the fermion-linear terms in $\delta_Q$ without the extra terms. Thus, all the higher-order terms cancel themselves desirably, leaving only field strengths.[5] There are many other intrinsic consistency checks, such as the $SO(2)$-invariance of the Chern-Simons terms in the lagrangian, in particular with $\widetilde{G}_3$, whose details will be reported elsewhere [17].

It is noteworthy that the scalar potential is positive definite:

$$V = +\tfrac{1}{32}\, e^{8\varphi/\sqrt{7}} \left[ 2(S_{ij})^2 - S^2 \right] = +\tfrac{1}{32}\, e^{8\varphi/\sqrt{7}} \left[ (A-C)^2 + 4B^2 \right] \quad , \tag{2.13}$$

agreeing with [10]. Note that this positive definite potential for gauged maximal supergravity is very peculiar to 9D. The main technical reason is that the index range is $i = 1, 2$, so that even if all the $\sigma$-model scalar fields have zero backgrounds, the potential (2.13) is minimized to zero by the balance between $2(S_{ij})^2$ and $S^2$. This feature is rather uncommon to other gauged maximal supergravity theories in other dimensions [8]. For example, even though the gauged maximal supergravity in 8D [15] has formally the same combination $2(T_{ij})^2 - T^2$, this scalar potential is not positive definite in 8D, due to the different index range $i = 1, 2, 3$ for the coset $SL(3, \mathbb{R})/SO(3)$. This situation is similar to other lower dimensions, such as gauged $N = 4$ supergravity in 7D [16], or $N = 8$ gauged supergravity in 4D [18] all of which have scalar potentials with regions with negative values.

[4] We cite [9] with the caveat mentioned in the Introduction in mind.

[5] We have confirmed this fact explicitly up to quintic order terms. Our result also agrees with [10] up to notation-dependent numerical coefficients.

## 3. 7-Brane Domain Solutions and Supersymmetry

We next analyze the stability analysis of 7-brane domain wall solutions. Even though there have been general formulae developed, based on Bogomol'nyi equations [19] for domain wall solutions in arbitrary space-time dimensions [20][21], we still need to look into the explicit forms of Killing spinor conditions for surviving supersymmetry. There are a few reasons for this.

First, for example, it was claimed in [11] that when all the $\sigma$-model scalar fields have zero backgrounds, minimizing the scalar potential, this vacuum has no preserved supersymmetry. However, this conclusion disagrees with [10]. It is also rather unusual that the background solution minimizing the scalar potential lacks supersymmetry. This is because in an ordinary supergravity theory, it is more common and natural that the minimal point of the scalar potential maintains supersymmetry. We analyze this problem of supersymmetry for $(\mathrm{Mink})_9$, by keeping all the three mass parameters $m_1$, $m_2$ and $m_3$.

Second, general prescriptions based on Bogomol'nyi equations in [20][21] seem to overlook certain subtleties related to the internal indices and $\gamma$-matrix properties depending on different space-time dimensions. For example, eq. (12) in [20] with the $\gamma$-matrices as a general form common to any space-time dimensions, needs some care. In fact, since ref. [20] uses the signature $(-++\cdots+)$, all the $\Gamma_m$-matrix in eq. (12) in [20] should be replaced by $i\Gamma_m$ in the case of 9D, as is clear from our explicit transformation rule (2.6) based on the signature $(+--\cdots-)$. The existence of such an imaginary unit causes a crucial difference for the Killing spinor equations. Due to these subtleties, we look into the Killing spinor equations directly instead of using the general prescriptions [20][21].

We start with the space-time metric of 7-brane domain wall solutions

$$ds^2 = e^{2\alpha(r)}\,\eta_{\underline{\mu}\underline{\nu}}\,dx^{\underline{\mu}}\,dx^{\underline{\nu}} - dr^2 \quad , \tag{3.1}$$

where $r \equiv x^8$, and $\underline{\mu},\,\underline{\nu},\,\cdots = 1,\,2,\,\cdots,\,7$, so that $(\eta_{\underline{\mu}\underline{\nu}}) \equiv \mathrm{diag.}\,(+\overbrace{--\cdots-}^{7})$, and the last term is negative due to $g_{88} = -1$. Now the three Killing spinor equations are $\delta_Q\psi_\mu = 0$, $\delta_Q\chi_i = 0$, $\delta_Q\lambda = 0$, where all the scalars $\alpha$, $\phi$, $\rho$ and $\varphi$ depend only on $r$. Under these conditions, and other bosonic fields with zero backgrounds, we get the Killing spinor conditions

$$\phi' = \pm\tfrac{1}{2} g e^{4\varphi/\sqrt{7}}(A - C) \quad , \tag{3.2a}$$

$$\rho' = \pm g e^{4\varphi/\sqrt{7}-\phi} B \quad , \tag{3.2b}$$

$$\varphi' = \pm\tfrac{1}{2\sqrt{7}} g e^{4\varphi/\sqrt{7}}(A + C) = -2\sqrt{7}\alpha' \quad , \tag{3.2c}$$

$$\alpha' = \mp\tfrac{1}{28} g e^{4\varphi/\sqrt{7}}(A + C) \quad , \tag{3.2d}$$

where the prime $'$ denotes the derivative with respect to $r$, and $A$, $B$, $C$ are given in (2.4). As has been mentioned, these forms are 'roughly' the same as those derived from general Bogomol'nyi equations in [20][21]. The important difference is that the Dirac matrix $\gamma^{(8)}$ in our signature $(+-\cdots-)$ has no real eigenvalues due to $(\gamma^{[8]})^2 = -I$, so that a condition like $\gamma^{[8]}\epsilon = \pm\epsilon$ does not make sense. The key point is the necessity of $\tau_3$ in front, and the right chirality on 8D for the Killing spinor $\epsilon$ should be like $\tau_3\gamma^{(8)}\epsilon = \pm\epsilon$.[6] Accordingly, the right condition for a general Killing spinor equation

$$(a + b\tau_3\gamma^{(8)})\epsilon = 0 \quad , \tag{3.3}$$

with generally field-dependent coefficients $a$ and $b$ is

$$a \pm b = 0 \quad , \tag{3.4}$$

respectively for $\tau_3\gamma^{(8)}\epsilon = \pm\epsilon$ in our signature $(+--\cdots-)$. In fact, all the $\gamma^{(8)}$ in all the Killing spinor equations in (2.6) for the domain wall metric appear always with $\tau_3$ in front. Note also that we do not use the arbitrary constant $c$ [10] in so-called 'superpotential' $W$ [21][20], because our computation based on Killing spinor equations necessitate *no* such an arbitrary constant $c$.[7]

Our field equations for the bosonic fields $\phi,\ \rho,\ \varphi,\ \alpha$ and $A_\mu$ are obtained as

$$\phi'' + 8\alpha'\phi' - e^{2\phi}(\rho')^2 - \tfrac{1}{4}g^2 e^{8\varphi/\sqrt{7}}(A^2 - C^2) \doteq 0 \quad , \tag{3.5a}$$

$$\rho'' + 8\alpha'\rho' + 2\phi'\rho' - \tfrac{1}{2}g^2 e^{8\varphi/\sqrt{7}-\phi}B(A + C) \doteq 0 \quad , \tag{3.5b}$$

$$\varphi'' + 8\alpha'\varphi' - \tfrac{1}{4\sqrt{7}}g^2 e^{8\varphi/\sqrt{7}}\Big[\,(A - C)^2 + 4B^2\,\Big] \doteq 0 \quad , \tag{3.5c}$$

$$\alpha'' + \tfrac{1}{14}(\phi')^2 + \tfrac{1}{14}e^{2\phi}(\rho')^2 + \tfrac{2}{7}(\varphi')^2 \doteq 0 \quad , \tag{3.5d}$$

$$(\alpha')^2 - \tfrac{1}{112}(\phi')^2 - \tfrac{1}{112}e^{2\phi}(\rho')^2 - \tfrac{1}{28}(\varphi')^2 + \tfrac{1}{14}V \doteq 0 \quad , \tag{3.5e}$$

$$2B\phi' - e^{\phi}(A - C)\rho' \doteq 0 \quad , \tag{3.5f}$$

where $\doteq$ denotes a field equation. When $m_1 = m_2 = 0$, (3.5) and (3.2) agree with the corresponding equations in ref. [10].

It is not too difficult to show that the Killing spinor condition (Bogomol'nyi equation) (3.2) is the sufficient condition of all the bosonic field equations in (3.5). This provides us with a good consistency check of the total system. Relevantly, since the condition (3.2) is stronger than the field equations (3.5), there are certain solutions that satisfy field equations (3.5), but not the former. This is nothing but the well-known fact that certain background solutions do not maintain supersymmetry.

As the most basic case, consider first the background solutions $\phi = \rho = \varphi = \alpha = 0$. Obviously, this can not satisfy (3.2), because $A = C = B = 0$ leads to $m_0 = m_1 = m_2 = 0$ which means no gauging. However, it is true that this set of solutions trivially satisfy all the bosonic field equations (3.5) including the $A_\mu$-field equation. To avoid misunderstanding, we emphasize that the vacuum solution is *not* realized, unless $m_0 = m_1 = m_2 = 0$ or $g = 0$. In other words, if $g = 0$ with no gauging from the outset, there *is* a supersymmetric $(\text{Mink})_9$ as the trivial background, in agreement with [11].

[6] This is because $\tau_3\gamma^{[8]}$ has real eigenvalues. When in the opposite signature $\big(-++\cdots+\big)$, $\gamma^{(8)}$ should have an imaginary unit in front, but the presence of $\tau_3$ is still essential for Killing spinor equations.

[7] In our notation, so-called 'superpotential' [10][21][20] is proportional to $W \approx S = \eta_{\alpha\beta}g^{\alpha\beta}$.

As for the puzzle that the minimizing point of the scalar potential does not preserve supersymmetry, we take the flowing standpoint. Namely, this $(\text{Mink})_9$ solution represents only the vacuum but not domain walls, so that the usual argument based on the Nester tensor [22][23] does not apply. To be more specific, the Nester tensor [22] in our system

$$N^{\mu\nu} \equiv (\overline{\eta}\gamma^{\mu\nu\rho}\widehat{\mathcal{D}}_\rho\eta) \tag{3.6}$$

satisfies the off-shell identity

$$D_\nu N^{\mu\nu} = (\widehat{\mathcal{D}}_\nu\overline{\eta})\gamma^{\mu\nu\rho}(\widehat{\mathcal{D}}_\rho\eta) + (\delta_\eta\overline{\chi}_i)\gamma^\mu(\delta_\eta\chi_i) + (\delta_\eta\overline{\lambda})\gamma^\mu(\delta_\eta\lambda) - 2e^{-1}(\overline{\eta}\gamma^\nu\eta)\,\frac{\delta\mathcal{L}}{\delta g^{\mu\nu}} \quad , \tag{3.7}$$

where $\eta$ is a commuting spinor for supersymmetry, distinguished from the anti-commuting one $\epsilon$, while $\delta_\eta$ is $\delta_Q$ with $\epsilon$ replaced by $\eta$. Accordingly, $\widehat{\mathcal{D}}_\mu\eta$ is the same as $\delta_Q\psi_\mu$ with $\epsilon$ replaced by $\eta$. The last term in (3.7) disappears on-shell. The $D_\mu N^{\mu 0}$ is negative definite under the Witten condition $\gamma^\mu\delta_\eta\psi_\mu = 0$ [22][23], so that the integration of $N^{\mu\nu}$ over a space-like boundary that encloses the domain wall is positive definite [22][23]

$$\tfrac{1}{2}\int_{\partial\Sigma} d\Sigma_{\mu\nu}\, N^{\mu\nu} = \int_{\partial\Sigma} d\Sigma_{0\nu}\, N^{0\nu} = -\int_{\Sigma} d\Sigma_0\, D_\mu N^{\mu 0} \geq 0 \quad . \tag{3.8}$$

On the other hand, the surface integral $\int d\Sigma_{0\mu}N^{0\mu}$ can be separately evaluated with two terms coming from $\delta_Q\psi_\mu$ (2.6b): The tension (energy density) $\sigma$ of the domain wall and the central charge term with $gS$ [23]. Now the main obstruction against $(\text{Mink})_9$ to preserve supersymmetry in the gauged case is eq. (3.2c) or (3.2d) from $\delta_Q\psi_\mu = 0$ not satisfied, while (3.2a) and (3.2b) are just the same as the minimization of the scalar potential $V$. This is equivalent to state that $\widehat{\mathcal{D}}_\mu\eta$ is *not* zero, when the potential is minimized, unless trivially $g = 0$ or $m_0 = m_1 = m_2 = 0$. Moreover, in the case of $(\text{Mink})_9$, there is no domain wall from the outset, so that there is no domain wall tension $\sigma$ in (3.7). Therefore the usual argument based on the domain wall and AdS background [23] does not apply here, either. In any case, using Nester tensor argument turns out to be equivalent to analyzing the Killing spinor conditions.

The second basic case is when $\varphi \neq 0$, $\rho = \phi = 0$. First, the Killing spinor equations (3.2a) and (3.2b) imply that $A = C$, $B = 0$, which in turn leads to $m_1 = m_2 = 0$, $A = C = m_0 \equiv m$. Now (3.2c), with the upper sign without loss of generality, implies that

$$\varphi' = \frac{1}{\sqrt{7}}\, gme^{4\varphi/\sqrt{7}} \quad , \tag{3.9}$$

which can be easily integrated to give

$$e^{-4\varphi/\sqrt{7}} = -\tfrac{4}{7}gmr + \tilde{b}_0 \quad , \qquad e^{2\alpha} = \mu e^{-\varphi/\sqrt{7}} = (a_0 gmr + b_0)^{1/4} \quad , \tag{3.10}$$

with certain constants $a_0$ , $b_0$, $\tilde{b}_0$, $\mu$. The satisfaction of the Killing spinor conditions guarantees the stability as well as the satisfaction of all the bosonic field equations. This is similar to [10], except that we do not have such an ambiguity as the constant $c$ in [10].

A more interesting case is when $\rho = 0$, $\phi \neq 0$ , $\varphi \neq 0$ similar to [10]. In this case, (3.2) becomes

$$
\begin{aligned}
\varphi' &= +\tfrac{1}{\sqrt{7}} g e^{4\varphi/\sqrt{7}} (m_0 \cosh\phi + m_2 \sinh\phi) = -2\sqrt{7}\alpha' \quad , \\
\phi' &= +g e^{4\varphi/\sqrt{7}} (m_0 \sinh\phi + m_2 \cosh\phi) \quad .
\end{aligned}
\tag{3.11}
$$

Dividing these two equations, we can integrate over $r$ to get

$$
m_0 \sinh\phi + m_2 \cosh\phi = \tilde{\mu} e^{\sqrt{7}\varphi} \quad , \tag{3.12}
$$

where $\tilde{\mu}$ is a real constant. This can be used in (3.11) to get the integration

$$
\int \frac{d\phi}{(m_0 \sinh\phi + m_2 \cosh\phi)^{11/7}} = \tilde{a} g r + \tilde{b} \quad , \tag{3.13}
$$

with some real constants $\tilde{a}$ and $\tilde{b}$. This integral can be performed as

$$
\begin{aligned}
\int d\phi\ (p e^{\phi} + q e^{-\phi})^{\nu} &= \sum_{n=0}^{\infty} \frac{\Gamma(\nu+1)\, p^n\, q^{\nu-n}}{n!\, \Gamma(\nu - n + 1)\, (2n - \nu)} e^{(2n-\nu)\phi} + \text{const.} \\
&= -\frac{1}{\nu} q^{\nu} e^{-\nu\phi} F\left(-\nu, -\tfrac{\nu}{2}, \tfrac{2-\nu}{2}; -\tfrac{p}{q}\, e^{2\phi}\right) + \text{const.}
\end{aligned}
\tag{3.14}
$$

The first equality is from the expansion: $(P+Q)^{\nu} = \sum_{n=0}^{\infty} \Gamma(\nu+1)/[\, n!\, \Gamma(\nu-n+1)]\, P^n\, Q^{\nu-n}$, which is further simplified by the standard hypergeometric function [24],

$$
F(\alpha, \beta, \gamma; z) \equiv \frac{\Gamma(\gamma)}{\Gamma(\alpha)\, \Gamma(\beta)} \sum_{n=0}^{\infty} \frac{\Gamma(\alpha+n)\, \Gamma(\beta+n)}{n!\, \Gamma(\gamma+n)} z^n \quad . \tag{3.15}
$$

Therefore, the integral in (3.13) is (3.14) with $p \equiv (m_2 + m_0)/2$, $q \equiv (m_2 - m_0)/2$ and $\nu = -11/7$. This solution is a generalization of the domain wall solution in [10] to the case $m_2 \neq 0$. The only caveat here is that in [10] the special value $c = -1$ was taken for the arbitrary constant in so-called 'superpotential' $W$, while we have *no* such an ambiguity within our direct construction, as has been demonstrated.

To summarize, our the 7-brane domain wall solution is

$$
\begin{aligned}
& e^{11\phi/7} F\left(\frac{11}{7}, \frac{11}{14}, \frac{25}{14}; \tfrac{m_0+m_2}{m_0-m_2}\, e^{2\phi}\right) = a g r + b \quad , \\
& e^{2\alpha} = \hat{\mu} e^{-\varphi/\sqrt{7}} = \mu (m_0 \sinh\phi + m_2 \cosh\phi)^{-1/7} \quad ,
\end{aligned}
\tag{3.16}
$$

where $a$, $b$, $\mu$, $\hat{\mu}$ are appropriate real constants. This solution preserves a half of the original $N = 2$ supersymmetry, guaranteeing its stability. This is also consistent with the general expectation based on Nester tensor [22][23]. Our solution is a generalization of a similar solution in [10] to the case of $m_1 \neq 0,\ m_2 \neq 0$ with the caveat that we do not use the constant $c$ in [10].

Even though we do not exhaust all the cases of other non-trivial scalar fields in order to save space in this Letter, our results for the field equations (3.5) and Killing spinor equations (3.2) with the total lagrangian with the general mass parameters $m_0$, $m_1$ and $m_2$ will be of great help for possible future studies.

## 4. Concluding Remarks

In this paper we have performed a direct construction of gauged $N = 2$ supergravity in 9D, with the lagrangian (2.5) and the supersymmetry transformation rule (2.6). We did not use the dimensional reductions [7][5] from 11D or 10D, due to several advantages. First, we can directly confirm the consistency of the whole system, performing the variation under supersymmetry. Since such a confirmation is indispensable even for dimensional reductions [7][5], it is more economical to commence with the construction directly within 9D. In fact, it is by such a direct computation that we can exclude practical mistakes like sign-errors or overlooked matrix $\sigma_2$ which are fatal for supersymmetry. Second, when dealing only with bosonic fields in dimensional reduction [7][5], there is always some ambiguity for field redefinitions. However, it is the fermionic sector that plays a decisive role for the total consistency *via* supersymmetric invariance. Moreover, since the fermionic transformation rules are crucial for Killing spinor equations, we can not discard fermionic contributions, as well as the total lagrangian at the same time; They should be treated simultaneously with mutual consistency. In fact, in our direct construction, we have no ambiguity such as the constant $c$ given in [10] in so-called 'superpotential' $W$ [20][21]. Third, there may be some new degrees of freedom for the direct construction that are not easily obtained by dimensional reductions, *e.g.,* similarly to gauged $N = 8$ supergravity within 4D [25]. Fourth, since there have been already certain results by dimensional reductions from 10D or 11D [9][10][11], the direct construction provides a new methodology, as well as the consistency check for the past results. Our results with the three mass parameters also provide more general couplings than those studied in the past.

There seem to be a few reasons for the 'delay' in the past for such a direct construction of gauged $N = 2$ supergravity within 9D. First, even though it sounds in principle very straightforward to obtain this system by a dimensional reduction [5] from 10D or 11D [8], a practical computation turns out to be unexpectedly complicated. While the purely bosonic sector has been studied with relatively detailed computations [6], it is the remaining fermionic terms with Noether/Pauli couplings that play a crucial role for Killing spinor equations as well as the consistency of the total action. Even though the number of lagrangian terms by dimensional reduction [7][5] looks rather 'small' as in (2.5), the invariance confirmation of the total action has a considerable number of sectors. Actually, there can be 63 different sectors of the structure $(\text{fermion}) \times (\text{boson}) \times (\text{boson}')$ at the cubic order in the variation of the lagrangian under supersymmetry even *before* gauging, aside from quartic fermion terms. Such an invariance confirmation does not seem to have been accomplished in [9], not to mention more basic linear-order closures on all the bosonic fields. Second, the $SO(2)$-gauging process needs additional 23 different sectors to be confirmed for the invariance, up to cubic fermionic terms in the variation. Third, as we have seen, the fermion $\chi_i$ has an extra $SO(2)$ index, that necessitates an additional irreducibility condition $\tau_i \chi_i = 0$. This simplifies some parts of the computation, but it also increases the number of different sectors in the variations.

We have accomplished in this Letter such complicated computations by a direct construction within 9D, and we have the total control of this system. We have also confirmed the non-trivial structures of field strengths, such as $F_{\mu\nu\rho\sigma}$ in (2.7). In addition to the lagrangian and transformation rule, we also clarified the Killing spinor (Bogomol'nyi) equations (3.2) and bosonic field equations (3.5) for the 7-brane domain wall configuration (3.1) with the mass parameters $m_0,\ m_1,\ m_2$. As an application of our most general result, we have given a new domain wall solution (3.16) as a generalization of a similar solution in [10]. Even though we have not exhausted all the other possible solutions, these equations constitute the working ground for possible future studies of domain wall solutions for the gauged $N = 2$ supergravity in 9D.

By analyzing gauged maximal supergravity in 9D, we have encountered some new aspects of supergravity. For example, the most fundamental $(\text{Mink})_9$ background does not maintain supersymmetry in the gauged case. This is peculiar to the 9D case, because ordinary gauged maximal supergravity [8] has negative cosmological constant, leading to AdS background, such as in 10D massive type IIA [1], gauged $N = 4$ supergravity in 7D [16], or gauged $N = 8$ supergravity in 5D [26]. Therefore, it seems unusual to have the scalar potential (2.13) manifestly positive definite with no cosmological constant at the minimum, without preserving supersymmetry in $(\text{Mink})_9$. However, we have understood this situation from the standpoint that the usual argument of stability [23] based on the Nester tensor [22] does not apply to such a vacuum solution, which does not have domain walls with boundaries.

Even though we have performed only the gauging of the $SO(2)$-subgroup of $SL(2,\mathbb{R})$, it will be straightforward to generalize our result to the non-compact $SO(1,1)$-gauging, following the prescription in [27], with the lagrangian and the transformation rule well established at hand.

We could also repeat similar direct constructions of maximal supergravity theories with more general mass parameter in dimensions $D \leq 8$, such as $N = 4$ supergravity in 7D [16] or $N = 8$ supergravity in 5D [26]. Even though these new mass parameters are naturally understood as the generalizations [5] of Scherk-Schwarz dimensional reductions [7], direct constructions have certain advantages from a practical viewpoint.

We are grateful for C.M. Hull, P. Meessen, and E. Sezgin for helpful discussions. Special acknowledgment is addressed to E. Bergshoeff for informing us about ref. [28], where domain wall solutions in $N = 2$ supergravity in 9D are systematically given.

## References

[1] L. Romans, Phys. Lett. **169B** (1986) 374.

[2] H. Singh, *'Duality Symmetric Massive Type II Theories in D=8 and D=6'*, hep-th/0109147, JHEP **0204** (2002) 017.

[3] E. Bergshoeff, C.M. Hull and T. Ortín, hep-th/9504081, Nucl. Phys. **B451** (1995) 547; E. Bergshoeff, M. de Roo, M.B. Green, G. Papadopoulos, P.K. Townsend, hep-th/9601150, Nucl. Phys. **B470** (1996) 113.

[4] L. Randall and R. Sundrum, hep-th/9905221, Phys. Rev. Lett. **83** (1999) 3370; *ibid.* **83** (1999) 4690; J. Maldacena, hep-th/9711200, Adv. Theor. Math. Phys. **2** (1998) 231; H.J. Boonstra, K. Skenderis and P.K. Townsend, hep-th/9807137, JHEP **9901** (1999) 003.

[5] I.V. Lavrinenko, H. Lü and C.N. Pope, hep-th/9710243, Class. Quant. Grav. **15** (1998) 2239.

[6] P. Meessen, T. Ortín, hep-th/9806120, Nucl. Phys. **B541** (1999) 195.

[7] J. Scherk and J.H. Schwarz, Nucl. Phys. **B153** (1979) 61.

[8] *'Supergravity in Diverse Dimensions'*, Vols. **1** & **2**, A. Salam and E. Sezgin, *eds.*, North-Holland, World Scientific (1989); *and references therein.*

[9] N. Khviengia and Z. Khviengia, *'D=9 Supergravity and p-Brane Solitons'*, hep-th/9703063.

[10] P.M. Cowdall, *'Novel Domain Wall and Minkowski Vacua of D=9 Maximal SO(2) Gauged Supergravity'*, hep-th/0009016.

[11] J. Gheerardyn and P. Meessen, hep-th/0111130, Phys. Lett. **B525** (2002) 322.

[12] C.M. Hull, *private communications.*

[13] S.J. Gates, Jr., H. Nishino and E. Sezgin, Class. and Quant. Gr. **3** (1986) 21.

[14] T. Kugo and P.K. Townsend, Nucl. Phys. **B211** (1983) 157.

[15] A. Salam and E. Sezgin, Nucl. Phys. **B258** (1985) 284.

[16] M. Pernici, K. Pilch and P. van Nieuwenhuizen, Phys. Lett. **143B** (1984) 103.

[17] H. Nishino and S. Rajpoot, *in preparation.*

[18] B. De Wit and H. Nicolai, Phys. Lett. **108B** (1982) 285.

[19] E.B. Bogomol'nyi, Sov. Jour. Nucl. Phys. **24** (1976) 449; E. Witten and D. Olive, Phys. Lett. **78B** (1978) 77; K. Shiraishi and S. Hirenzaki, Int. Jour. of Mod. Phys. **A6** (1991) 2635; E. Moreno, C. Nuñez and F.A. Schaposnik, hep-th/9802175, Phys. Rev. **D58** (1998) 025015; H. de Vega and F.A. Schaposnik, Phys. Rev. Lett. **56** (1986) 2564, Phys. Rev. **D34** (1986) 3206.

[20] G.W. Gibbons and N.D. Lambert, hep-th/0003197, Phys. Lett. **B488** (2000) 90.

[21] A. Chamblin and G.W. Gibbons, hep-th/9909130, Phys. Rev. Lett. **84** (2000) 1090.

[22] J.M. Nester, Phys. Lett. **A83** (1981) 241; E. Witten, Comm. Math. Phys. **80** (1981) 241; P.K. Townsend, Phys. Lett. **148B** (1984) 55.

[23] K. Behrndt and M. Cvetič, hep-th/9909058, Phys. Lett. **B475** (2000) 253; M. Cvetič, S. Griffies and S.J. Rey, hep-th/9201007, Nucl. Phys. **B381** (1992) 301; W. Boucher, Nucl. Phys. **B242** (1984) 282.

[24] I.S. Gradshteyn and I.M. Ryzhik, *'Table of Integrals Series and Products'*, Academic Press, New York (1980).

[25] B. de Wit and H. Nicolai, Phys. Lett. **108B** (1982) 285; C. Hull, Phys. Lett. **142B** (1984) 39.

[26] M. Günaydin, L.J. Romans and N.P. Warner, Phys. Lett. **154B** (1985) 268; Nucl. Phys. **B272** (1986) 598.

[27] C.M. Hull, *'Gauged D=9 Sueprgravities and Scherk-Schwarz Reduction'*, hep-th/0203146.

[28] E. Bergshoeff, U. Gran and D. Roest, hep-th/0203202, Class. Quant. Grav. **19** (2002) 4207.